\documentstyle[preprint,aps,epsf]{revtex}

\begin{document}
\draft

\title{Nearest pattern interaction and global pattern formation}
	
\author{Seong-Ok Jeong, Hie-Tae Moon and Tae-Wook Ko}
\address{Department of Physics, Korea Advanced Institute 
	of Science and Technology, Taejon 305-701, Korea}
\date{\today}
\maketitle
\begin{abstract}

We studied the effect of nearest pattern interaction on a global pattern formation in a 2-dimensional space, where patterns are to grow initially from a noise in the presence of a periodic supply of energy. Although our approach is general, we found that this study is relevant in particular to the pattern formation on a periodically vibrated granular layer, as it gives a unified perspective of the experimentally observed pattern dynamics such as oscillon and stripe formations, skew-varicose and crossroll instabilities, and also a kink formation and decoration.

\end{abstract}
\narrowtext

Patterns arising in parametrically forced extended media such as Faraday systems, 
vibrated granular layers, and electrically driven liquid crystals 
have drawn interests in recent years\cite{aaa}.
From a dynamic point of view, it is noteworthy that recent experimental studies of the patterns arising in a vertically vibrated thin granular layer have led to the discovery of the so-called oscillons, a fluidlike localized nonlinear wave\cite{nat,oscil}. Following the discovery of oscillons, patterns with different dynamic characteristics in a periodically vibrated granular layer were also reported. These are (i) the patterns featuring the skew-varicose and crossroll instabilities in a fluid convection\cite{skew} and (ii) the kinks; the interfaces or fronts that separate domains of different values of phases\cite{oscil,das}. 

An understanding of the fundamental nonlinear dynamical behavior of 
granular materials remains a serious challenge\cite{rmp1,rmp2}.
Several theoretical approaches, including molecular dynamics simulations, order parameter equations, and hydrodynamic-type models, have been proposed to describe this phenomenology. Direct molecular dynamics simulations\cite{das,bizon} reproduced a majority of patterns observed in experiments and many features of the bifurcation diagram, although until now have not yielded oscillons and interfaces.
Hydrodynamic and phenomenological models\cite{mo1,roth,ott} reproduced certain experimental features; however, neither of them offered a systematic description 
of the whole rich variety of the observed phenomena\cite{blair}. As it is, it is often argued that a general local, continuum description of granular media, analogous to hydrodynamics, is unlikely to exist. 

Earlier, we proposed a nearest pattern interaction model to capture the essential phenomenology of the birth of oscillons. As this model was proposed before the experimental reports of the other types of patterns (i) and (ii) mentioned above, it remains primitive. In this study we would like to generalize this model to capture also the dynamic phenomenology of the patterns (i) and (ii).   

We start by presenting and briefly outlining the nearest pattern interaction model which we proposed earlier to study oscillons. For more details, readers may refer to the Ref.\cite{pre}. 

\begin{equation}
{\bf M}~:~ h_{n+1}(\vec{r})=F[h_{n}(\vec{r})+ \alpha G( \overline{ \Delta h_{n}(\vec{r})})],~~ n=1,2,....
\end{equation}

with

\begin{equation}
\overline{ \Delta h_{n}(\vec{r})} \equiv \frac{1}{\Delta S} 
\int W_{\vec{r'} \vec{r}} 
[h_{n}(\vec{r'})-h_{n}(\vec{r})] d^{2}\vec{r'}
\end{equation}
,where $\vec{r'}$ runs over entire space but $W_{\vec{r'}\vec{r}}$ is a 
weighting function being $1$ if $R \le |\vec{r'} -\vec{r}| \le 3R$ but 0 
otherwise, and $\Delta S$ denote the area of the concentric region where $W=1$.
We set $h_{n}(\vec{r'})$ to be coupled with its neighbors in terms of the introduced
averaged fluctuation $\overline{\Delta h_{n}(\vec{r})}$, which effectively
allows the interaction only with its nearest patterns (refer to Fig. 1(a)). This model employs a discrete mapping for time evolution. For a periodically forced oscillator, the discrete time step often corresponds to the period of forcing.

We assume that patterns are to grow from a noise, subject to a statistical constraint that a rise($h>0$) or a fall($h<0$) from the reference flat surface($h=0$) is equally likely over time. Such constraint satisfies the mass conservation. 
To implement such assumption in the model, we took the following odd function as our functional form of $F$ (See Fig.1(b)).

\begin{equation}
F(h)=\left\{\begin{array}{ll}
     ~h     & \textrm{for}~ |h| \le h_{0},  \\
     ~h_{0} & \textrm{for} ~ h>h_{0},  \\
     -h_{0} & \textrm{for} ~ h<-h_{0}.
     \end{array} \right.
\end{equation}

With such functional form of $F$, the noise does not grow and remains as a fixed point if their is no nonlinear coupling with neighbors, i.e., if $G=0$. 

For the following specific form of $G$ (See Fig. 1(c)), we learned in our earlier study that the model is capable of producing the oscillons, and by varying the coupling constant $\alpha$, the model could capture the formation of stripes, oscillon chains and oscillon lattice, etc\cite{pre}.

\begin{equation}
G(\Delta h)=tan \frac{\pi}{2} \Delta h/(2h_{0})
\end{equation}

The significance associated with other forms of $F$ and $G$ was a question in our earlier study which we could not answer in the absence of other forms of patterns (i) and (ii). With more patterns available now, we may proceed to generalize the model as follows.

Let's start with the function $G$. The function $G$ is meant to incorporate
the interaction between neighboring excitations,
so we can change the nonlinearity of the 
interaction by using another function for $G$.
Earlier, we had chosen highly nonlinear function Eq.(4) for $G$,
to give sufficient hysteresis for stable oscillon structures.
What if we choose a less nonlinear function for $G$?
In such a case, the model exhibits skew-varicose and crossroll instabilities, while the stability region of oscillons becomes narrower. For example, using a simple linear function $G(\Delta h)=\Delta h$, and setting $\alpha=1.5$,
we obtain the pattern dynamics as shown in Fig. 2. 
We have checked many forms of $G$ such as piecewise linear functions and simple nonlinear polynomials. 
In these cases, we obtain in general the patterns as featured in Fig. 2 as long as $G$ is odd and monotonically increasing. However, these instabilities do not occur if $G$ is highly nonlinear as in the case of Eq.(4), 
for the stability boundary becomes too wide.
We note that the stability of convection rolls in a fluid heated from below is
limited by these secondary instabilities.
Recently, an experimental study has shown
that such pattern dynamics are also found in granular systems\cite{skew}.

Next, we turn to the role of the function $F$. We notice immediately that by 
changing the slope of the function $F$, we can make the system a bistable system. 
For that purpose, we introduce a new parameter $\beta$ as follows.

\begin{equation}
F(h)=\left\{\begin{array}{ll}
     ~\beta h     & \textrm{for}~ \beta |h| \le h_{0},  \\
     ~h_{0} & \textrm{for} ~\beta h>h_{0},  \\
     -h_{0} & \textrm{for} ~\beta h<-h_{0}.
     \end{array} \right.
\end{equation}

For $\beta>1$, $h=0$ becomes a unstable equilibrium, while 
$h= +h_{0}$ and $h= -h_{0}$ become stable equilibria(see Fig. 3.), making the 
system bistable.
Conceivably, as there are now two stable attractors separated by an unstable one, the domain may be divided into two parts, one with high elevation corresponding to $h= +h_{0}$ and the other with low depletion corresponding to $h= -h_{0}$, 
separated by a phase front, namely, a kink.
The model now contains two control parameters, $\alpha, \beta$, which have 
a definite role respectively. That is, $\alpha$ measures the degree of nonlinear 
coupling with the neighbors, leading into self organization, and $\beta$ 
determines the degree of nonequilibriumness of the system. 

For $\beta>1$, and for the tangent function of Eq. (4) for $G$, the results are
summarized in Fig. 4. When $\alpha$ is
below a certain critical value, $\alpha_{c1}$, indicated by the filled circles 
in Fig. 4, the map synchronizes $h$'s with their neighbors, so flats
occur. In this case, however, there now emerge two possible phases of flats 
which are separated by an interface
called a ``kink" \cite{oscil}.
Figure 5(a) shows a typical kink solution of our model. 
As we increase $\alpha$ above $\alpha_{c1}$, the interfaces are decorated
by periodic undulations as shown in Fig. 5(b,c).
We found that the areas embedded within the region of the opposite phase
as in Fig. 5(b) are unstable and shrinks, while fronts which extend to the
wall(as in Fig. 5(c)) are stable.
We point out that the undulated kinks shown in Fig. 5(b),(c) resemble the the decorated kinks reported recently in shaken granular beds\cite{oscil,das}.
In our model, decorated parts oscillate at $f=1/2$ while
flats are $f=1$. But in experiments, both flats and decorations are oscillating
at $f=1/2$. This is because, in real experiments, flat regions undergo 
a period doubling while a front separating these regions does not\cite{oscil}.

Associated with the kinks, the present model further shows the followings. First, as $\alpha$ exceeds $\alpha_{c2}$, 
corresponding to the region denoted by the filled squares in 
Fig. 4, the decorated kinks begin to be elongated to generate stripes. 
Fig. 6(a) shows the decorations elongating
into stripes as $\alpha$ is raised to the stripe region($\alpha = 16$).
The same phenomenology is observed in experiments\cite{das,aa2}.
Second, the decoration can be converted into
oscillon chains when we decrease $\beta$ below unity, as in Fig. 6(b,c).
This point implies that oscillon chains can emerge from the decorated kinks when vibrational acceleration is decreased abruptly to the oscillon region, and we have confirmed that this phenomenology is indeed observed in a laboratory experiment\cite{aa2}. 
This point is noteworthy because the undulation in our model is not
merely a simple curvature of interfaces, but related with oscillons\cite{cmt}. This aspect is similar to the stripes turning into a chain of oscillons as conformed experimentally.

In conclusion, we have constructed a dynamical model to understand the 
key mechanism for the pattern formation in a parametrically forced spatially extended medium. 
The nearest pattern interaction model here contains two control parameters $\alpha$ and $\beta$; 
$\beta$ controls the 
number of stable equilibrium states of the system, and $\alpha$ monitors
the nearest pattern interaction.  
Despite its extreme mathematical simplicity, 
our model turns out to provide a unified perspective of the various patterns 
observed experimentally in periodically vibrated media.
 
The authors thank H. K. Pak for useful discussions and providing valuable
experimental informations. This work was supported in part by the
interdisciplinary research program Grant No. 1999-2-112-002-5 of KOSEF and in part by BK21 program of Ministry of Education in Korea.

\pagebreak
\newpage
{\large \bf Figure Captions}
\begin{description}
\item[Fig. 1.] (a) A nearest pattern interaction approximation. The field at $P$
is coupled with the fields at other locations only within the shaded region.
Here R represents the domain size of an excitation. The functions (b) $F$ and
(c) $G$ used in Ref.\cite{pre} are presented.
\item[Fig. 2.] Secondary instabilities; (a) Skew-varicose instability;
(b) Crossroll instability.
\item[Fig. 3.] The function $F$ ($\beta > 1$). 
The slope $\beta$, which changes the local
dynamics of $h$, influences the overall pattern dynamics drastically.
See the text for details.
\item[Fig. 4.] Phase diagram in general. The open circles and squares
are to indicate the region of hysteresis. Conditions for numerical
simulations are the same with Ref.\cite{pre}. See the text for details.
\item[Fig. 5.] Various types of kinks; (a) linear kinks ($\alpha=8.0,
\beta=1.5$); (b,c) Decorated
kinks ($\alpha=13.0, \beta=1.5$).
\item[Fig. 6.] (a) Kinks under kink to stripe transition ($\alpha=16.0,
\beta=1.5$). (b,c) Oscillon chains from decorated kinks in Fig. 4(b,c)
($\alpha=13.0, \beta=0.95$).

\end{description}

\begin{references}
\bibitem{cmt} Here, the emerging oscillons are not super oscillons which are reported in \cite{blair}, but conventional ones. 
It is also noteworthy that stripes, not labyrinths, are formed by the
elongation of decorated kinks.
\end{references}
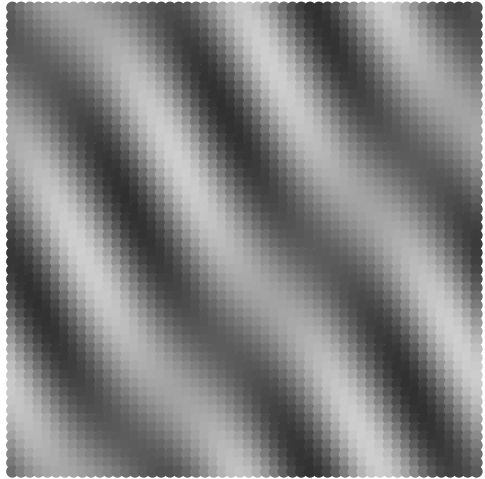
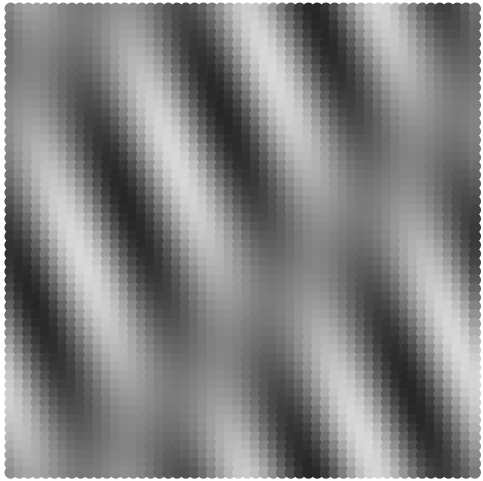
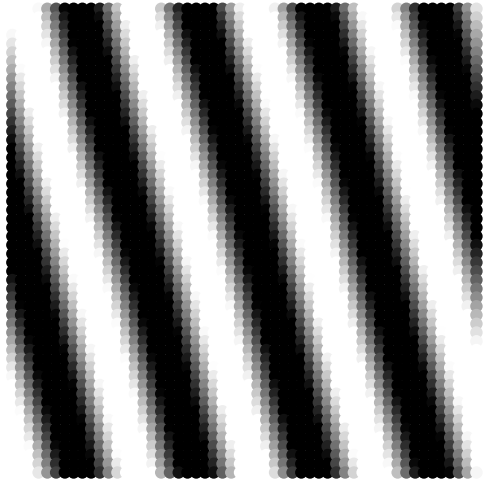
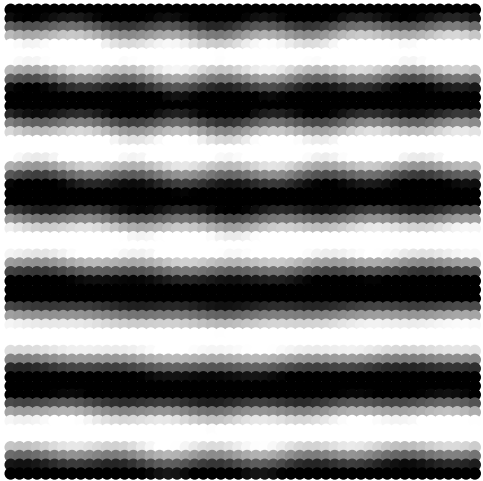
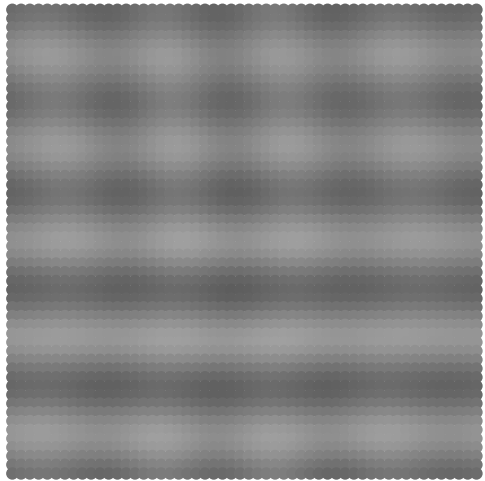
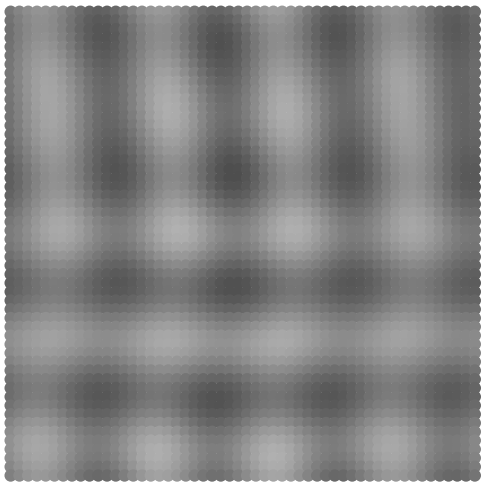
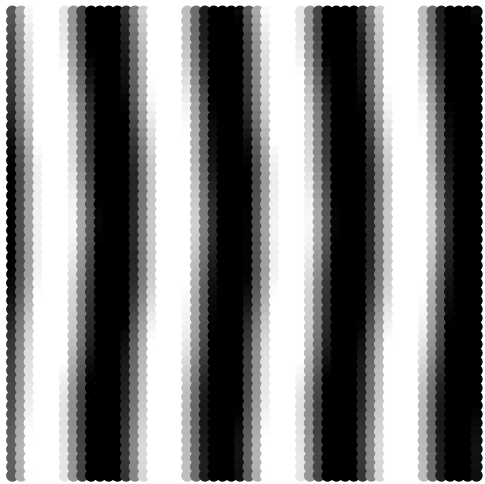
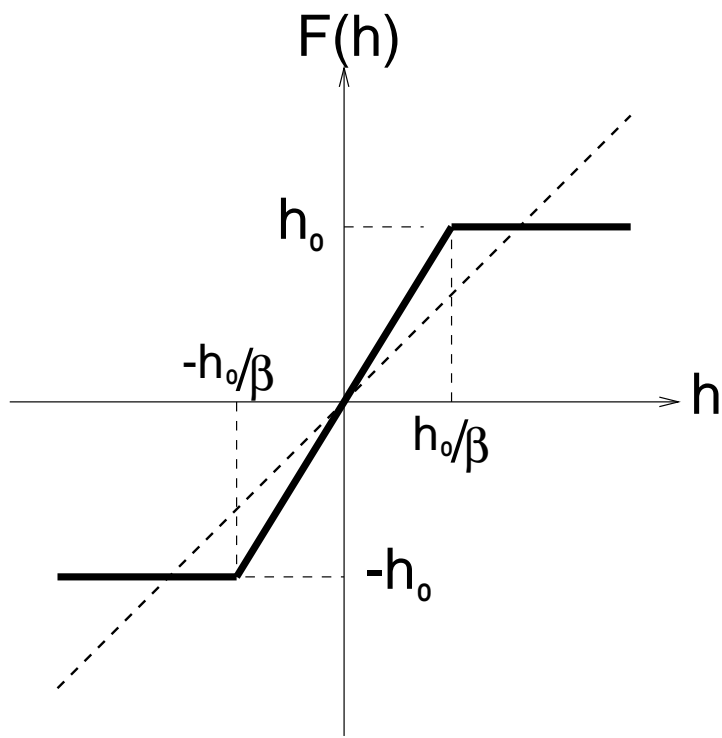
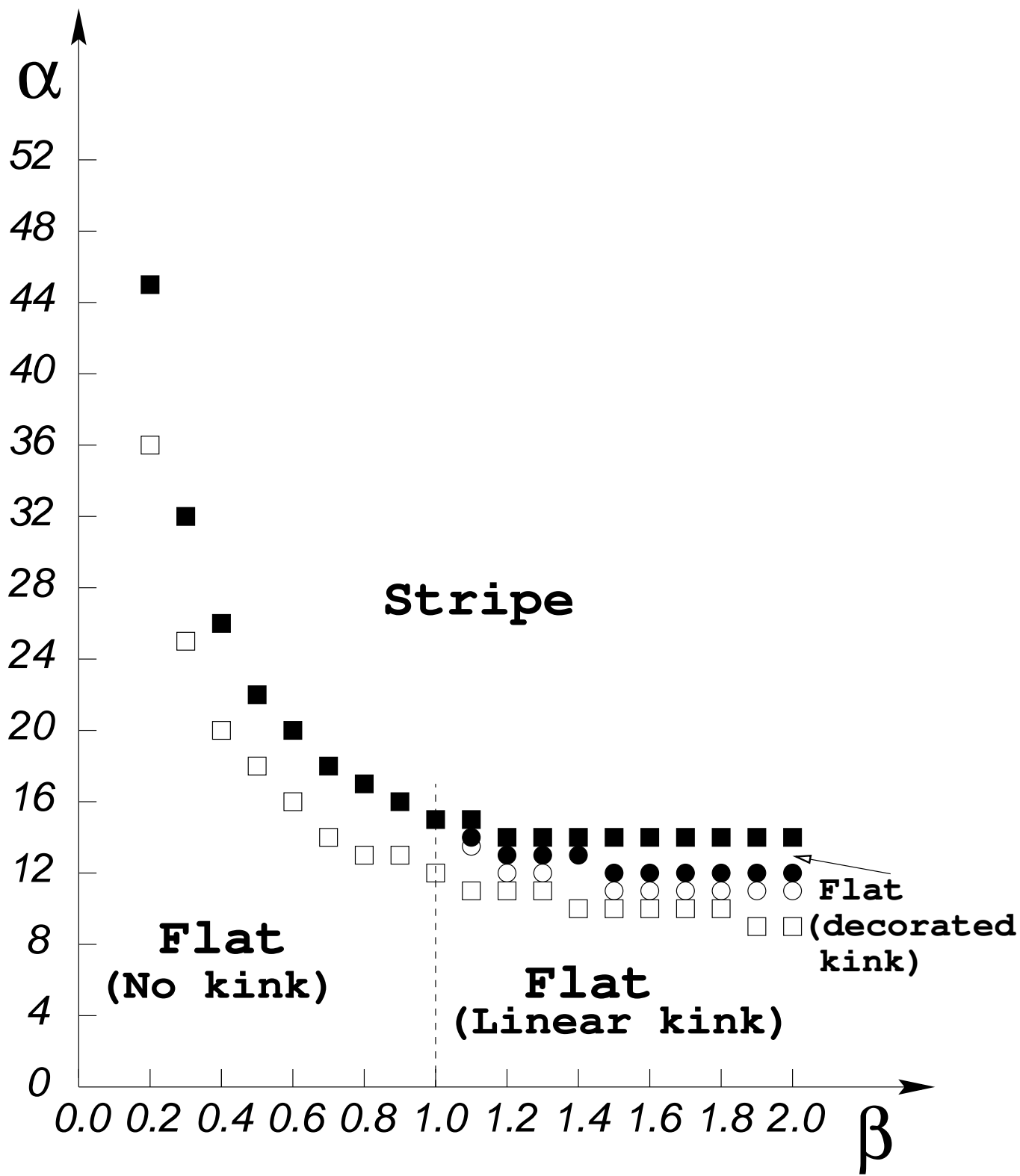
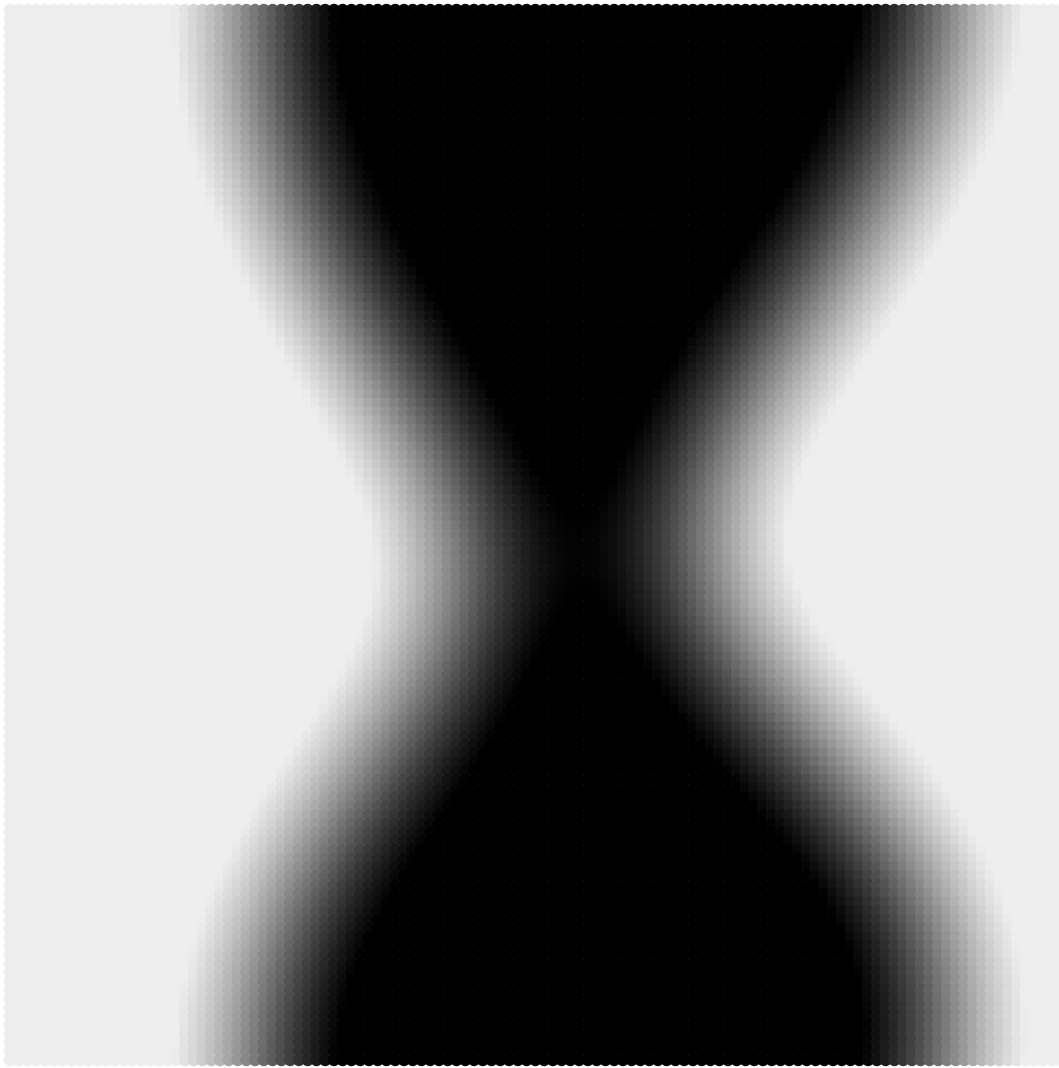
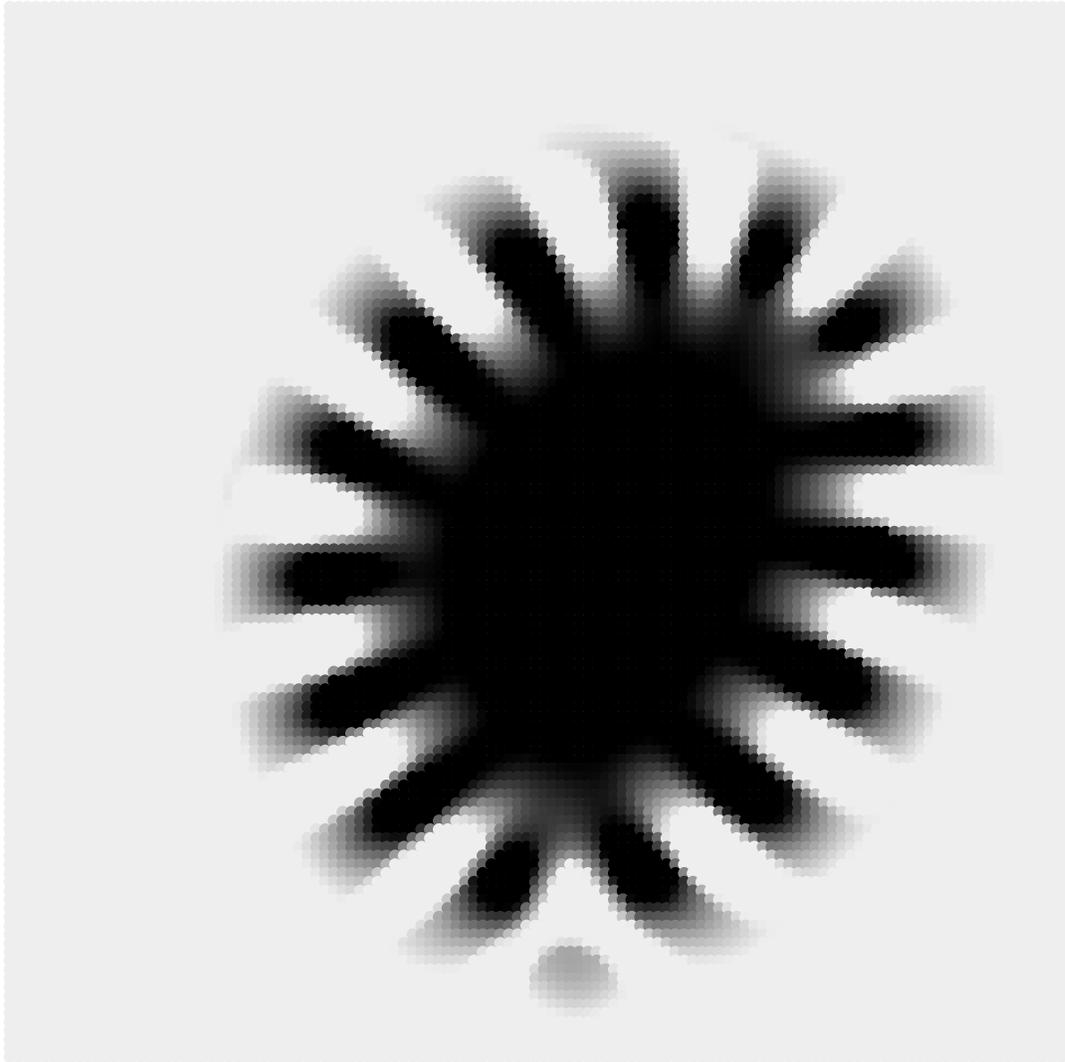
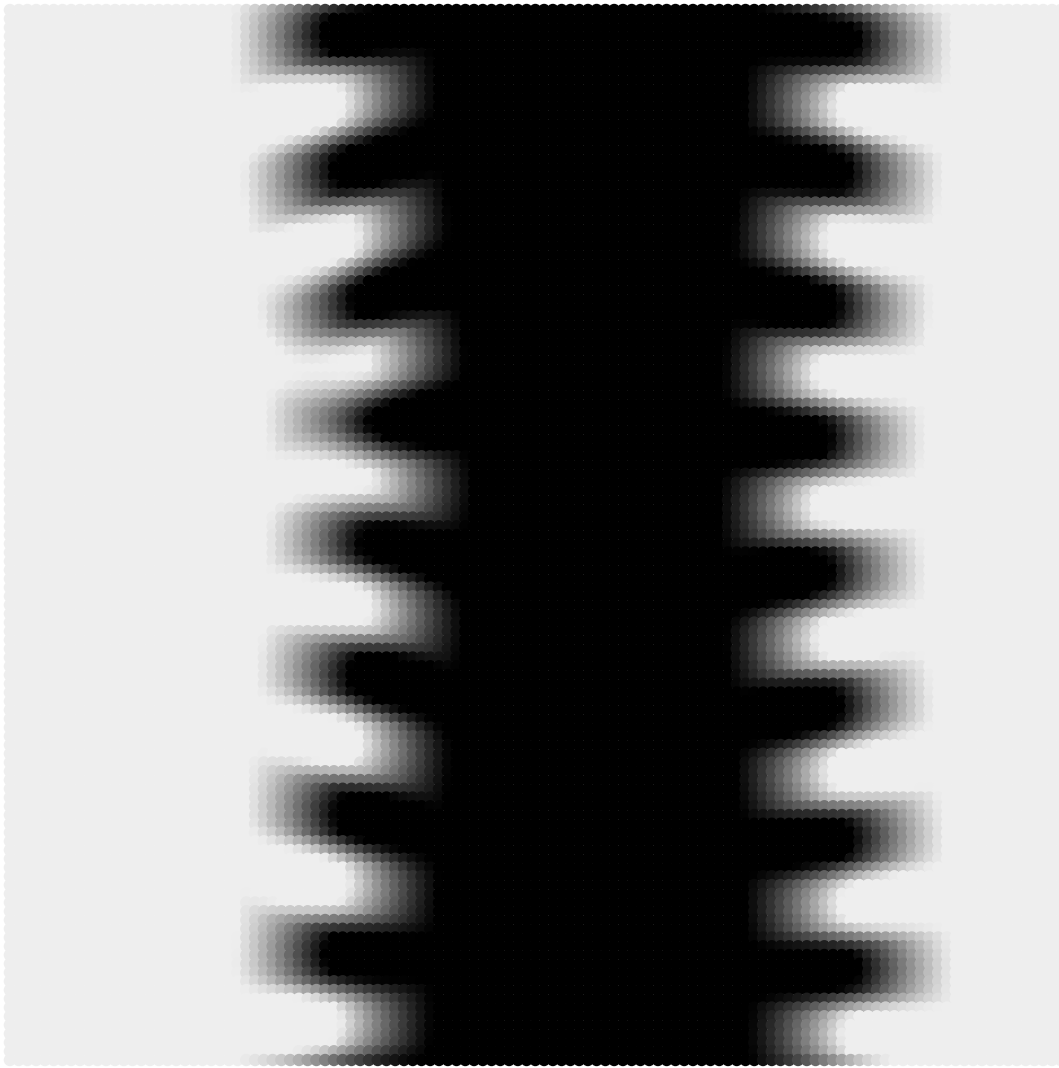
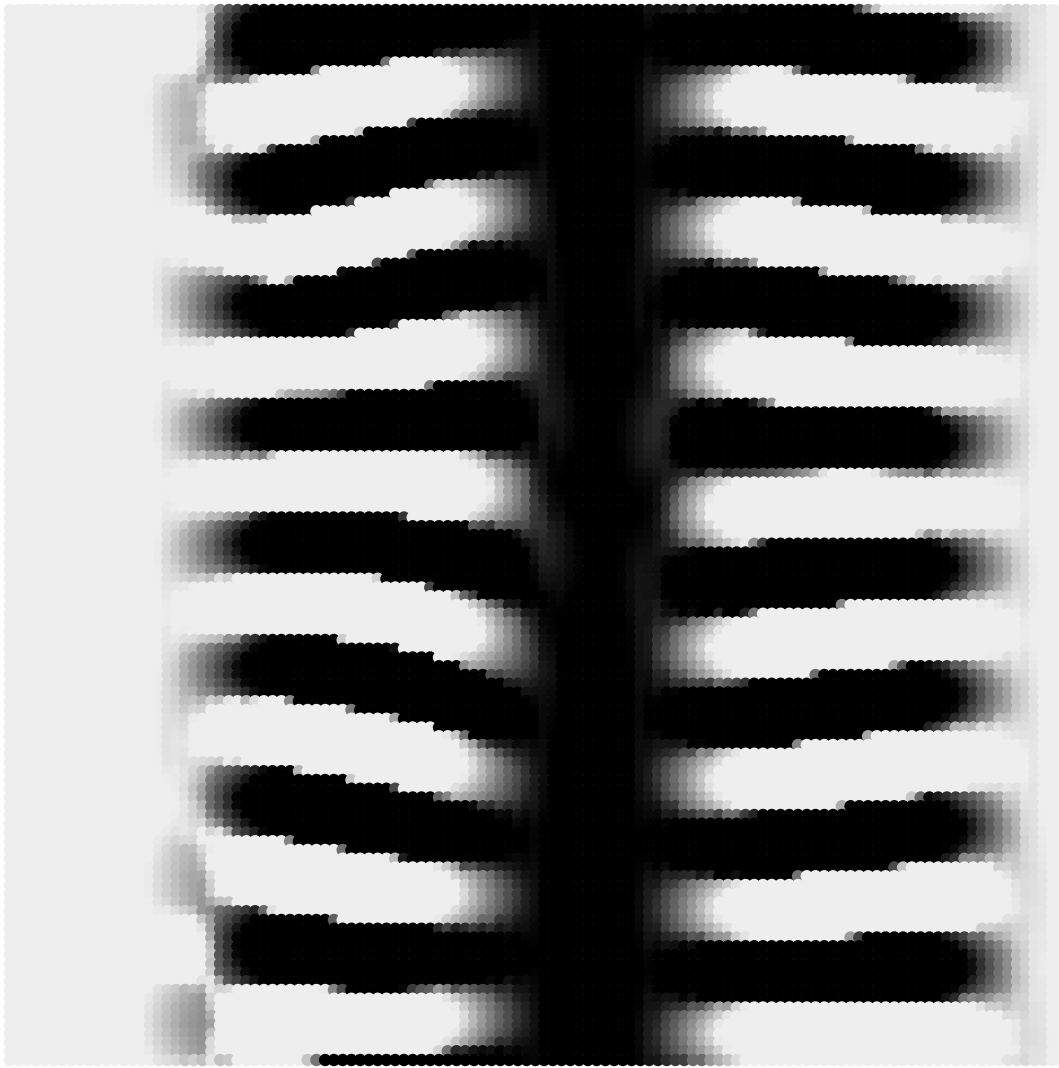
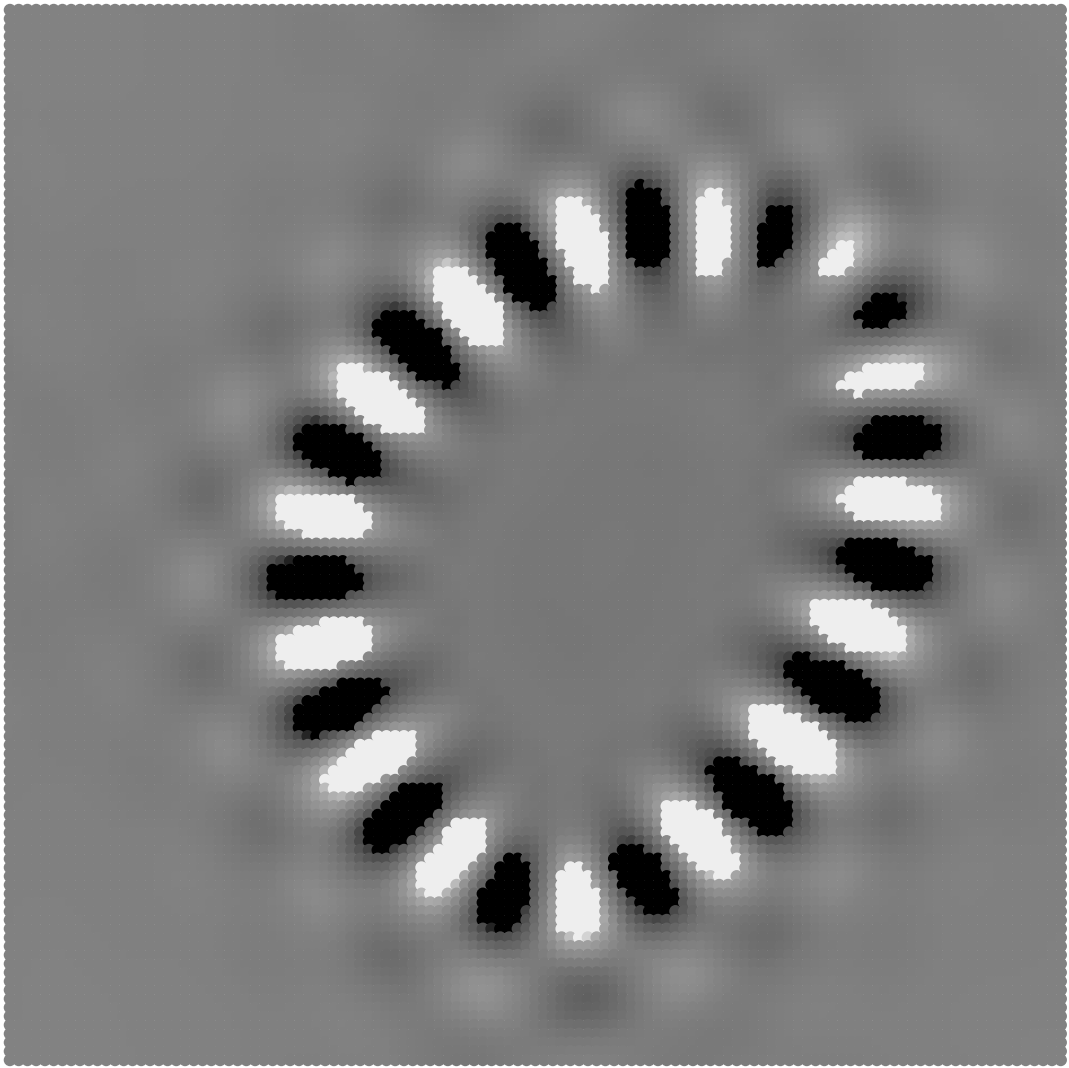
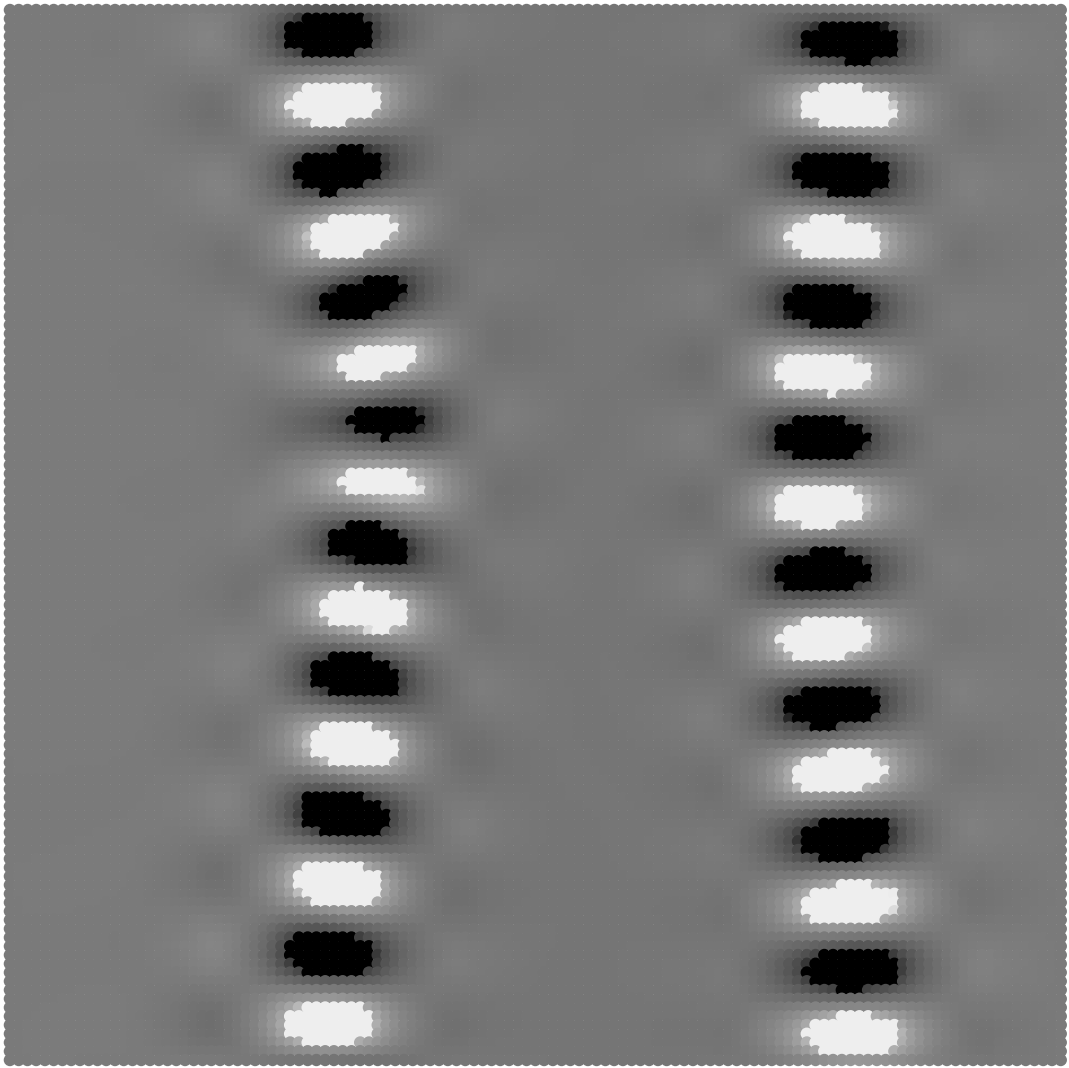
\end{document}